# GENERALIZED SIMULATION MODEL OF TEACHING AND ITS RESEARCH ON PC

Mayer Robert, *Glazov State Pedagogical Institute, Glazov*

One of the important problems of cyber pedagogy is the following: how, knowing the parameters of the student, his initial level of knowledge and the impact of the teacher to predict knowledge of student at subsequent times. Simulation method allows you to create a computer program that simulates the behavior of the "teacher–student"–system and investigate the influence of system parameters on the results of learning.

Imagine students, each of which is characterized by a set of parameters $\alpha_i$, $\beta_i$, $\gamma_i$ ... ($i = 1, 2, ..., n$) and $m$ teachers who own methods $M_1$, $M_2$, $M_3$ etc. The main objective of didactics is such: so to organize the teaching process (to select methods and to distribute learning material for a predetermined period of time) that in the end of training students coped with the test $T = \{T_1, T_2, ...\}$. Let $U$ denote the level of teacher requirements. We formulate **the law of didactics: the rate of increase of knowledge $Z$ is proportional to student's efforts $F(t)$, efficiency of using training methods $k$, coefficient of assimilation $\alpha$ (learning coefficient) and coefficient of understanding $P$: $dZ/dt = aPkF(t)$.** We can assume that efforts of pupil $F(t)$ proportional to the difference between the level of teacher requirements and student knowledge $U - Z$.

Student's efforts $F(t)$ are characterize the intensity of his mental activity and the level of his motivation $M$. The contradiction between the level of requirements and the level of student's knowledge is the driving force behind its training activities. The more a student knows, the easier it sets associative links and more quickly assimilate new knowledge. Motivation to learn depends on the impact of the teacher or level teacher requirements $U$ (amount of knowledge which must be learned), and the level of knowledge $Z$ student. We do not distinguish between extrinsic motivation due to the requirements of the teachers, and intrinsic motivation, caused by desire of student to learn the appropriate discipline. The more $M$ (and hence $F$) and the level of student's knowledge $Z$, the easier it sets associative links and quickly assimilate new knowledge. Psychologists have found that if the level of the entry requirements is low, then there is no motivation to learn. If the requirements are too high and exceeds a critical value $C$, then it also reduces motivation. Ideally, the teacher should prop-erly assess the condition of the student and use such methods, in which his motivation is maximized. This mode of learning in which the requirements $U$ in optimal manner correspond to the level of knowledge of the student's $Z$ and his motivation is maximal we will call **coordinated.**

We will create **a simulation model of the learning process** [1 – 8]. The speed of increase $Z$ is proportional to learning coefficient $\alpha$, operability (workability) $r$, applied efforts $F$ and the level of knowledge $Z$ in degree b ($0 \leq b \leq 1$):

$$dZ/dt = r\alpha F Z^b - \gamma Z,$$

wherein $r$ — factor of operability which depend on the degree of fatigue student and $\gamma$ – forgetting factor. Value $r$ is first equal to $r_0$ ($0 < r_0 \leq 1$), and then gradually decreases to 0 according to the law: $r = r_0/(1+\exp(k_1(P-P_0)))$. Here $P_0$ — work done by the student (number of problem solving, performing tasks), in which his operability is reduced from $r_0 = 1$ to $r = 0{,}5$. On the lesson as rule the requirement of teacher is higher than the level of knowledge the student ($U > Z$), and instructional work done by the student (number of completed tasks), depends on the student's efforts (the intensity of mental activity) and the duration of training. Student's efforts $F_i$ for $\Delta t_i$ is proportional to their motivation or the difference between the level of $U_i$ the entry requirements and the amount of knowledge $Z_i$:

$$F_i = U_i - Z_i, \quad \Delta P_i = k_2 F_i \Delta t_i = k_2(U_i - Z_i)\Delta t_i, \quad P = k_2 \sum_{i=1}^{N} F_i \Delta t_i.$$

If the level of the teacher's requirements is small ($U \leq Z$), we have a student in class is busy solving simple problems for him. In this case student's efforts proportional to the time: $P = kt$. This allows to take into account the tiredness of the learner and reduction of his workability even in the case where it performs a simple task for a long time. In between lessons (on breaks) student are resting, his factor of operability is restored exponentially:

$$dr/dt = k_3(r_{\max} - r), \quad r(t) = r_{\max} - (r_{\max} - r_0)\exp(-k_3(t - t_0)),$$

where $r_0 = r(t_0)$ — factor of operability at the start of the break $t_0$, $r_{\max}$ — maximum workability of the student at a given time t of the school day. It gradually decreases on law $r_{\max} = \exp(-k_4 t)$. If subjective complexity (difficulty of understanding) $S$ of the studied material is small then the rate of increase of knowledge, all other things being equal, is higher. Therefore: $dZ/dt = r(1-S)\alpha F Z^b - \gamma Z$.
Complexity educational material $S$ is in the range from 0 to 1, and generally depends on the study of other issues. So, we have the model:

During training $(U > Z)$: $$\frac{dZ}{dt} = \frac{(1-S)\alpha F Z^b}{1+\exp(k_1(P-P_0))} - \gamma Z,$$

During the break $(U = 0)$: $dZ/dt = -\gamma Z$.

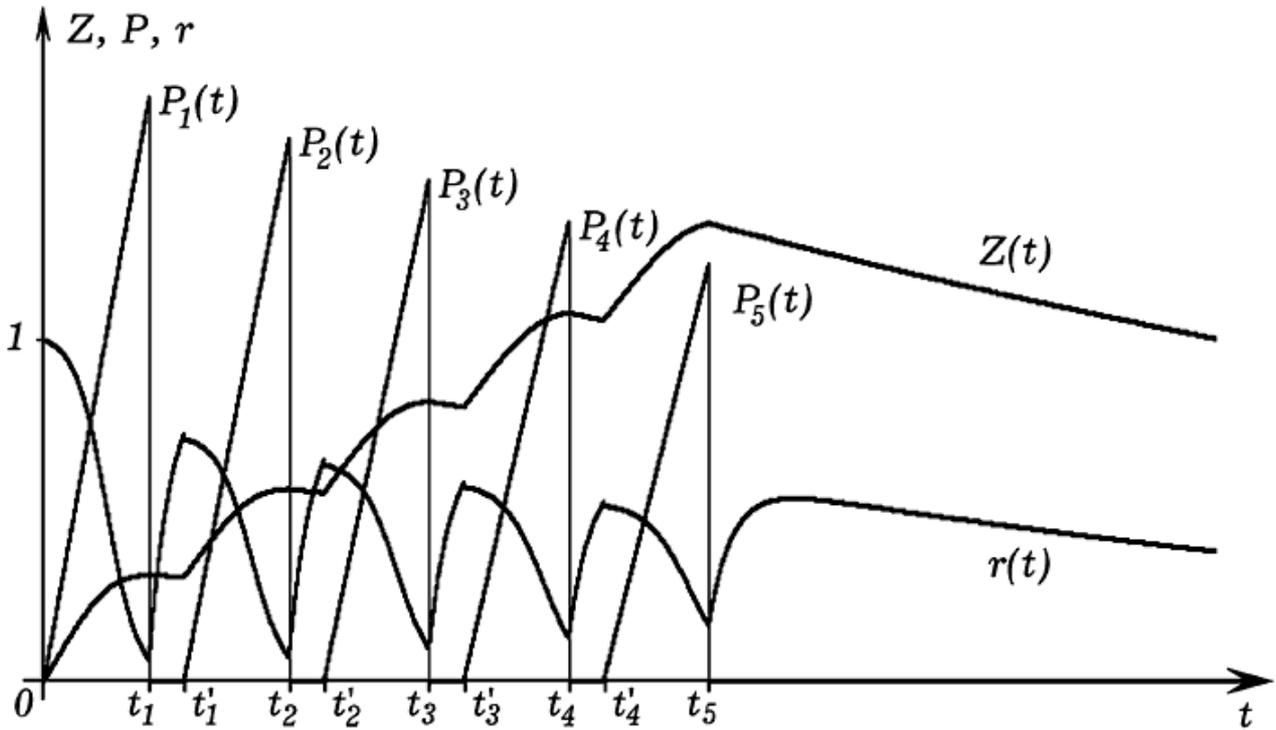

Fig. 1. The results of the simulation.

Let's say a teacher organizes the learning process so during the day, that students work with a maximum strain $F = const$. Increase of knowledge significantly less than the total number of student knowledge $Z$, therefore $\beta = 0$. Let five lessons held the same duration $T_u = t_1 = t_2 - t'_1 = ... = t_5 - t'_4$, separated by breaks lasting $T_p = t'_1 - t_1 = t'_2 - t_2 = ... = t'_4 - t_4$. Simulation results represented on Fig. 1. The model parameters are chosen so that its behavior is correspond with the actual situation. Coefficient of workability of student is oscillating relatively smoothly decreasing value in the range from 0 to $t_5$. When reducing the length of the breaks between classes, students do not have time to restore its value, so the results of training are worsen.

Complicate the model. When memorizing information is appearing associative links between new and existing knowledge. As a result the received knowledge becomes stronger and it is forgotten much slower. Let $Z$ — overall knowledge of the student, $Z_1$ — the most "weak" knowledge of the first category with a high rate of forgetting $\gamma_1$, $Z_2$ — knowledge of the second category with a lower coefficient of forgetting $\gamma_2$, ..., and $Z_n$ — the most "trustworthy" knowledge of the $n$–th category with the lowest $\gamma_n$ ($\gamma_1 > \gamma_2 > ... > \gamma_n$). Coefficient of assimilation (learning) $\alpha_i$ characterize the speed of knowledge transformation $(i-1)$–th category in a more solid knowledge of the $i$–th category. While a learning, $o=1$, and when it is ended, $o=0$. Forgetting factor $\gamma=1/\tau$, where $\tau$ – time, in which knowledge reduction in $e=2,72...$ times. Result of learning is characterized not only by the amount of acquired knowledge $Z = Z_1 + Z_2 + ... + Z_n$, but also the coefficient of "robustness" or "trustworthy" $\Pr = (Z_2/2^{n-2} + ... + Z_{n-1}/2 + Z_n)/Z$.

```
uses crt, graph; const a1=0.06; a2=0.002; g1=0.001;    { ПР - 1 }
g2=5E-5; dt=0.01; Tu=300; Tp=100; Mt=5; Mz=4; Mr=200;
var o,DV,MV : integer; F,Z,Z1,Z2,S,P,t,r,r0: real;
BEGIN DV:=Detect; InitGraph(DV,MV,'c:\bp\bgi');     {Free Pascal}
line(0,470,640,470); F:=3; r0:=1; Z:=0;
Repeat t:=t+dt; If t<Tu then O:=1; If t>Tu then O:=0; If t>Tu+Tp
then begin O:=1; S:=0; end; If t>2*Tu+Tp then begin O:=0; end;
If t>2*Tu+2*Tp then begin O:=1; S:=0.2; end; If t>3*Tu+2*Tp then
begin O:=0; end; If t>3*Tu+3*Tp then begin O:=1; S:=0.3; end;
If t>4*Tu+3*Tp then begin O:=0; end; If t>4*Tu+4*Tp then
begin O:=1; S:=0.4; end; If t>5*Tu+4*Tp then begin O:=0; end;
Z:=Z1+Z2; If O=1 then begin P:=P+0.2*(1+S)*F*dt;
r:=r0/(1+exp(0.03*(P-200))); Z1:=Z1+r*(1-S)*(a1*F-a2*Z1)*dt
-g1*Z1*dt; Z2:=Z2+a2*r*(1-S)*Z1*dt-g2*Z2*dt; end else begin
P:=0; r:=r+0.015*(exp(-2E-4*t)-r)*dt;
Z1:=Z1-g1*Z1*dt; Z2:=Z2-g2*Z2*dt; r0:=r; end;
circle(10+round(t/Mt),470-round(Mz*Z),1); circle(10+round(t/Mt),
470-round(Mz*Z2),1); circle(10+round(t/Mt),470-round(Mr*r),1);
circle(10+round(t/Mt),470-round(P*1.2),1);
until KeyPressed; CloseGraph;
END.
```

In the study one subject at first increases the level of knowledge $Z$, after that is increasing the share of "solid" knowledge $Z_n$ and coefficient of strength $\Pr$. The author proposed a generalized model of learning that is unparalleled in known to him literature. Let the initial student workability $r_0 = 1$.

At any time, $Z(t) = Z_1(t) + ... + Z_n(t)$.

During training ($o = 1$):

$$F = U - Z > 0, \quad r = r_0 / (1 + \exp(k_1(P - P_0))), \quad P = k_2 \int_{t_0}^{t} (1 + S)(U - Z)dt,$$

$$dZ_1 / dt = r(1 - S)(\alpha_1 F Z^b - \alpha_2 Z_1) - \gamma_1 Z_1,$$

$$dZ_2 / dt = r(1 - S)(\alpha_2 Z_1 - \alpha_3 Z_2) - \gamma_2 Z_2,$$

$$..., \quad dZ_n / dt = r(1 - S)\alpha_n Z_{n-1} - \gamma_n Z_n.$$

During the break ($o = 0$): $U = 0$, $dr/dt = k_3(r_{max} - r)$, $r_{max} = \exp(-k_4 t)$,

$$dZ_1/dt = -\gamma_1 Z_1, \quad dZ_2/dt = -\gamma_2 Z_2, \quad ..., \quad dZ_n/dt = -\gamma_n Z_n.$$

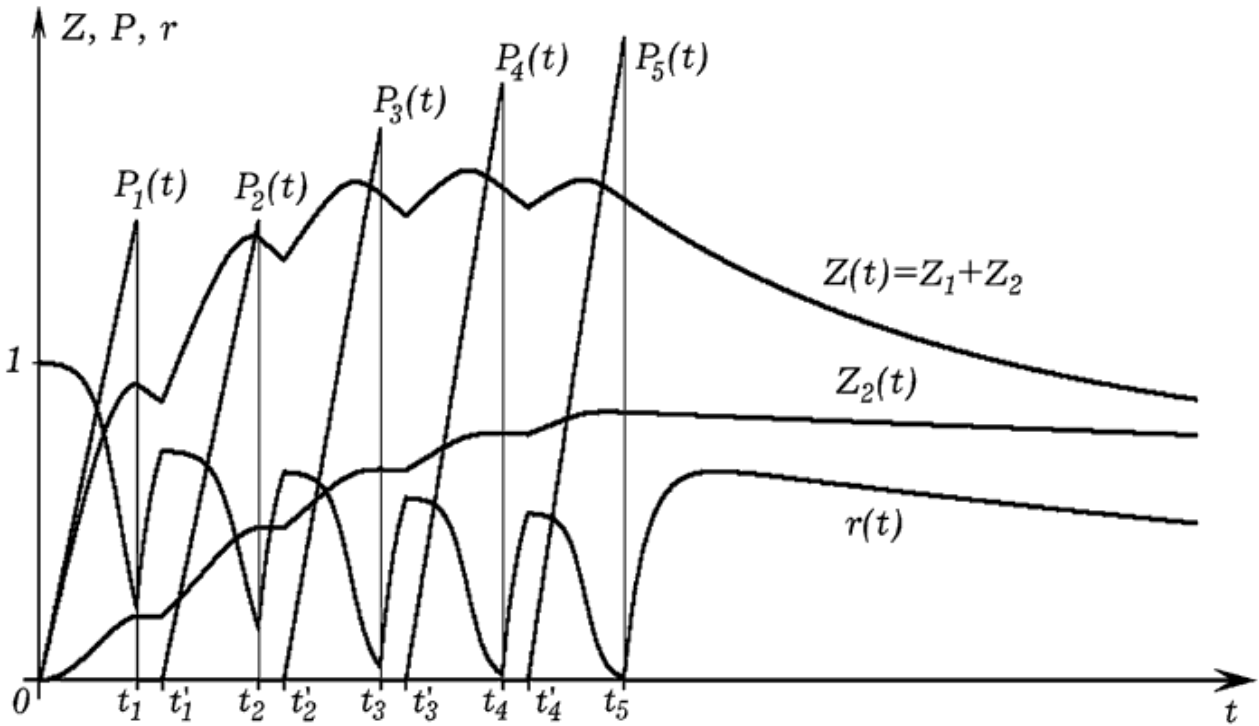

Figure 2. The simulation results (two-component model).

Simulation results is presented on Fig. 2 (program PR–1). Volume of "strong" (or "solid") knowledge Z2 in the learning process is growing, and after this knowledge not forgotten. Other models of learning can be found at http://rmajer.narod.ru (http://komp-model.narod.ru ).